\documentstyle[aps,prb]{revtex}
\begin{document}

\title{Order-parameter fluctuations in the frustrated Heisenberg model
on the square lattice}
\author{Shu Zhang and Gerhard M\"uller}
\address{Department of Physics, The University of Rhode Island, Kingston,
Rhode Island 02881-0817}
\date{\today}
\maketitle

\begin{abstract}
The $T=0$ dynamics of the two-dimensional $s=1/2$ Heisenberg model with
competing nearest-neighbor $(J_1)$ and next-nearest-neighbor $(J_2)$
interactions is explored via the recursion method, specifically
the frequency-dependent fluctuations of the order parameters associated
with some of the known or suspected ordering tendencies in this system,
i.e. N\'eel, collinear, dimer, and chiral order.
The results for the dynamic structure factors of the respective fluctuation
operators show a strong indication of collinear order at
$J_2/J_1 \gtrsim 0.6$ and a potential for dimer order at
$0.5 \lesssim J_2/J_1 \lesssim 0.6$, whereas the chiral ordering tendency
is observed to be considerably weaker.
\end{abstract}

\twocolumn

The impact of quantum fluctuations on the zero-temperature phase diagram
of a quantum many-body system tends to be strongest if that system
contains competing interactions.
A particular, sometimes exotic phase may be stabilized by quantum
fluctuations in the presence of interactions that frustrate each other's
ordering tendencies.
The two-dimensional (2D) Heisenberg antiferromagnet with nearest-neighbor
($nn$) and next-nearest-neighbor ($nnn$) coupling on the square lattice,
\begin{eqnarray}
H=&J&_1\sum_{\bf r}{\bf S}_{\bf r}\cdot({\bf S}_{\bf r+\hat{x}} +
                    {\bf S}_{\bf r+\hat{y}}) \nonumber \\
 +&J&_2\sum_{\bf r}{\bf S}_{\bf r}\cdot({\bf S}_{\bf r+\hat{x}+\hat{y}} +
                                {\bf S}_{\bf r+\hat{x}-\hat{y}}) \;,
\label{1}
\end{eqnarray}
has been a prominent object of study in this context.\cite{R93}

The N\'eel long-range order (LRO) present in the ground state of the
$nn$ model disappears at some critical coupling ratio,
$J_2/J_1 \simeq 0.4$, and is replaced, at sufficiently large $J_2/J_1$,
by N\'eel LRO within each of the two $nnn$ sublattices.
The latter is preceeded, at $J_2/J_1 \gtrsim 0.65$, by collinear
LRO, which breaks the (discrete) rotational symmetry of $H$
on the lattice but not yet its (continuous) rotational symmetry in
spin space.
The N\'eel and collinear order parameters (OPs) are described by
the operators,
\begin{equation}
O_N = {1 \over N} \sum_{\bf r}(-1)^{x+y}N_{\bf r} \;, \;\;\;\;
O_C = {1 \over N} \sum_{\bf r}C_{\bf r} \;,
\label{2}
\end{equation}
where $N_{\bf r} = S_{\bf r}^z$,
$C_{\bf r} = {\bf S}_{\bf r}\cdot({\bf S}_{\bf r + \hat{x}}
                                + {\bf S}_{\bf r - \hat{x}}
              - {\bf S}_{\bf r +\hat{y}} - {\bf S}_{\bf r - \hat{y}})$.

At intermediate coupling ratios, $0.4 \lesssim J_2/J_1 \lesssim 0.65$,
the N\'eel and collinear ordering tendencies keep each other at bay and
thus make the frustrated ground state susceptible to different kinds of
ordering potentialities.
Dimer order,\cite{GSH89CJ91,SN90} twist order,\cite{DM89} and
chiral order\cite{RGW9192} have been proposed in this context.
The dimer and chiral OPs considered here are
defined by the following expressions in terms of spin operators:
\begin{equation}
O_D = {1 \over N} \sum_{\bf r}(-1)^xD_{\bf r} \;,\;\;\;\;
O_\chi = {1 \over N} \sum_{\bf r}\chi_{\bf r}\;,
\label{3}
\end{equation}
where $D_{\bf r} = {\bf S}_{\bf r}\cdot{\bf S}_{\bf r + \hat{x}}$,
$\chi_{\bf r}$ = $Z_{\bf r,r+\hat{x},r+\hat{x}+\hat{y}}$ $-$
$Z_{\bf r,r+\hat{x}+\hat{y},r+\hat{y}}$,
$Z_{\bf i,j,k}$ = $i(S_{\bf i}^+S_{\bf j}^--%
S_{\bf i}^-S_{\bf j}^++S_{\bf j}^+S_{\bf k}^--%
S_{\bf j}^-S_{\bf k}^++S_{\bf k}^+S_{\bf i}^--%
S_{\bf k}^-S_{\bf i}^+)$.
They probe the long-range phase coherence of singlets stacked in columns
along the $y$-axis and the handedness of the spin configuration on a
plaquette, respectively, in the ground-state wave function
$|G\rangle$.
Whether any one of these types of LRO is, in fact, realized,
or whether a ground state with short-range correlations of the
resonating-valence-bond type, \cite{CD88,R90} for example, is
stabilized, has not been determined for certain.

The absence or presence of a specific type of LRO determines
whether the associated OP correlation function decays to
zero or not.
In the finite-cluster data, the relevant information on the asymptotic
behavior is best captured by the expectation value of the squared OP.
The problem here is to find a meaningful reference point for any
enhancement in that quantity.\cite{SN90}

An alternative avenue to comparing the different ordering tendencies
in the spin-frustrated ground state of (\ref{1}) is to explore the dynamic
(i.e. frequency-dependent) fluctuations of any  proposed OP.
The recursion method\cite{rm} in conjunction with recently developed
techniques of continued-fraction analysis\cite{VM94,VZSM94} is very
suitable for that purpose.
Here the dynamical information is derived from the finite-size
ground-state wave function.
No excited states have to be computed.
This is an important advantage for the study of systems with
complicated spectra and with potential OPs that have widely
varying symmetry properties, as is the case here.

We investigate the fluctuations of the four OPs defined
in (\ref{2}) and (\ref{3}) as they manifest themselves in the dynamic
structure factors
\begin{equation}
S_{AA}({\bf q},\omega) = \int\limits_{-\infty }^{+\infty }dt
e^{i\omega t}\langle F^A_{\bf q}(t)F^A_{\bf -q}\rangle \;,
\label{4}
\end{equation}
where $F^A_{\bf q} =
N^{-1/2} \Sigma_{\bf r}e^{i{\bf q}\cdot{\bf r}}A_{\bf r}$
is the fluctuation operator associated with a given OP,
and $A_{\bf r}$ stands for $N_{\bf r}$, $C_{\bf r}$, $D_{\bf r}$, or
$\chi_{\bf r}$, as defined above.

The recursion algorithm in the present context is based on an orthogonal
expansion of the wave function $|\Psi_{\bf q}^A(t)\rangle$ =
$F^A_{\bf q}(-t)|G\rangle$.
It produces (after some intermediate steps) a sequence of
continued-fraction coefficients $\Delta^A_1({\bf q}), \Delta^A_2({\bf q}),
\ldots $ for the relaxation function,
\begin{equation}
c_0^{AA}({\bf q},z) = \frac{1}{\displaystyle z + \frac{\Delta^{A}_1({\bf q})}
{\displaystyle
z +  \frac{\Delta^{A}_2({\bf q})}{\displaystyle z + \ldots }  }  } \;,
\label{5}
\end{equation}
which is the Laplace transform of the symmetrized correlation function
$\Re\langle F^A_{\bf q}(t)F^A_{-{\bf q}}\rangle/\langle
F^A_{\bf q}F^A_{-{\bf q}}\rangle$.
The $T=0$ dynamic structure factor is then obtained from (\ref{5}) via
\begin{equation}
S_{AA}({\bf q},\omega) =  4\langle F^A_{\bf q}F^A_{-{\bf q}}\rangle
\Theta(\omega)\lim \limits_{\varepsilon
\rightarrow 0} \Re [c_{0}^{AA} ({\bf q}, \varepsilon - i\omega)] \;,
\label{6}
\end{equation}
where$\langle F^A_{\bf q}F^A_{-{\bf q}}\rangle = S_{AA}({\bf q})$
is the integrated intensity.\cite{intensity}

All the results presented here are for clusters of
$N=4\times 4$ sites with periodic boundary conditions.
Extreme care must be exercised in separating finite-size effects from
properties that reflect the physics of the infinite system.
This distinction can be made with more confidence for coupling ratios
$J_2/J_1 \lesssim 0.7$ than at $J_2/J_1 > 0.7$, where the
gradual decoupling of the two $nnn$ sublattices causes a crossover in the
finite-size effects.

The N\'eel OP fluctuations are probed by the dynamic structure factor
$S_{NN}({\bf q},\omega)$ at the wave vector ${\bf q}=(\pi,\pi)$.
This quantity is shown in Fig. 1 for various coupling ratios.
The presence of N\'eel LRO at $J_2/J_1=0$ implies that
$S_{NN}(\pi,\pi,\omega)$ is governed by a zero-frequency peak.
Quantum fluctuations split the ground-state level for finite $N$.
In a $4\times 4$ cluster the finite-size gap is known to be of
magnitude $\Delta E/J_1 \simeq 0.57$.\cite{VZSM94}
The peak position of the curve for $J_2/J_1 = 0$ must be
interpreted with this fact in mind.

What are the effects of the $nnn$ coupling on the N\'eel OP
fluctuations?
For $0 \leq J_2/J_1 \lesssim 0.4$, i.e. over the estimated range of the
N\'eel phase, we observe only small changes in peak position and line
shape.
Then the N\'eel OP fluctuations begin to change rapidly in two stages:
(i) Over the range 0.4 $\lesssim$ $J_2/J_1$ $\lesssim$ 0.6, the peak
position moves to higher frequencies at an accelerated rate, the linewidth
shrinks, and the integrated intensity (not shown) drops to 32\% of its
value at $J_2/J_1=0$.
This signals the presence of some non-N\'eel type ordering tendency which
supports well-defined N\'eel modes at increasingly high frequencies.
(ii) At $J_2/J_1 \gtrsim 0.6$ the linewidth of $S_{NN}(\pi,\pi,\omega)$
grows rapidly, while the peak position moves further up and the integrated
intensity continues to fade away quickly.
As the two $nnn$ sublattices begin to decouple, the system
ceases to support well-defined N\'eel modes.

The dynamic structure factor $S_{CC}(0,0,\omega)$, which describes the
collinear OP fluctuations, is shown in Fig. 2.
At $J_2/J_1=0$, we observe a fairly sharp collinear mode at
$\omega/J_1 \simeq 3.0$.
As the N\'eel ordering tendency weakens with increasing $J_2/J_1$, the
collinear mode shifts to lower frequencies, while its line shape broadens
considerably.
The width reaches a maximum at $J_2/J_1 \simeq 0.4$.
Between here and $J_2/J_1 \simeq 0.55$, where the competing dimer and
chiral ordering tendencies are at their peak,
the collinear mode moves to $\omega=0$, and the integrated intensity
more than triples in relation to its value at $J_2/J_1=0$.
In the interval 0.55 $\lesssim$ $J_2/J_1$ $\lesssim$ 0.7,
the function $S_{CC}(0,0,\omega)$ transforms into a narrow central peak,
and $S_{CC}({\bf q})$ increases by another factor of $\simeq 1.6$.
This clearly reflects the onset of collinear LRO in the
infinite system.

The dimer OP fluctuations as described by the dynamic
structure factor $S_{DD}(\pi,0,\omega)$ and shown in Fig. 3 resemble those
of the collinear OP with respect to line shape and peak position
for as long as the N\'eel ordering tendency is perceptible in the ground
state $(J_2/J_1 \lesssim 0.4)$.
Both modes become soft and very broad at $J_2/J_1 \simeq 0.55$, but then
they part company.
While the collinear mode has been observed to transform into a high-intensity
narrow central peak, the dimer mode, which has reached its maximum
intensity here ($\simeq$ 1.8 times its value at $J_2/J_1=0$), broadens
further and and loses intensity very rapidly.
It literally dissolves as the inter-sublattice correlations begin to
weaken at $J_2/J_1 \gtrsim 0.7$.
Nevertheless, the softness of the dimer OP fluctuations at
$J_2/J_1 \simeq 0.55$ in the $4\times 4$ cluster is consistent with
dimer LRO in the infinite system.

The chiral OP fluctuations are based more significantly on intra-sublattice
correlations than the dimer OP fluctuations and, therefore, evolve
differently.
This is illustrated in Fig. 4.
At $J_2/J_1 =0$ the dynamic structure factor $S_{\chi\chi}(0,0,\omega)$
exhibits a sharp mode at $\omega/J_1 \simeq 3.8$.
At $J_2/J_1 \simeq 0.55$, the peak position has moved down to
$\omega/J_1 \simeq 1.6$, while the linewidth has increased only slightly,
and the integrated intensity has grown to a maximum value of $\simeq 2.3$
times its value at $J_2/J_1 = 0$.
Then the peak position starts to move back out to higher frequencies,
the line shape begins to broaden, but less so compared to that of the dimer
fluctuations, and the integrated intensity drops rapidly.
The minimum gap of the chiral mode is perhaps too large to be entirely
attributable to a finite-size effect,which would indicate that the
observed chiral ordering tendency does not turn into chiral LRO as
$N \rightarrow \infty$.

This work was supported by NSF Grant DMR-93-12252 and by the NCSA at
Urbana-Champaign.


\pagebreak

\begin{figure}
\caption[one]
{$T=0$ dynamic structure factor (\ref{4}) normalized by its integrated
intensity for the N\'eel fluctuation operator $F_{\bf q}^{N}$ at
${\bf q} = (\pi,\pi)$ of the Hamiltonian (\ref{1}) at various values
of the coupling ratio $J_2/J_1$, obtained via strong-coupling
continued-fraction reconstruction from the coefficients
$\Delta _1,...,\Delta _6$ and a Gaussian terminator as explained in Refs.
\onlinecite{VM94,VZSM94}.
}
\label{F1}
\end{figure}

\begin{figure}
\caption[two]
{$T=0$ dynamic structure factor (\ref{4}) normalized by its integrated
intensity for the collinear fluctuation operator $F_{\bf q}^{C}$ at
${\bf q} = (0,0)$ of the Hamiltonian (\ref{1}) at various values
of the coupling ratio $J_2/J_1$, obtained via strong-coupling
continued-fraction reconstruction from the coefficients
$\Delta _1,...,\Delta _6$ and a Gaussian terminator.}
\label{F2}
\end{figure}

\begin{figure}
\caption[three]
{$T=0$ dynamic structure factor (\ref{4}) normalized by its integrated
intensity for the dimer fluctuation operator $F_{\bf q}^{D}$ at
${\bf q} = (\pi,0)$ of the Hamiltonian (\ref{1}) at various values
of the coupling ratio $J_2/J_1$, obtained via strong-coupling
continued-fraction reconstruction from the coefficients
$\Delta _1,...,\Delta _6$ and a Gaussian terminator.}
\label{F3}
\end{figure}

\begin{figure}
\caption[four]
{$T=0$ dynamic structure factor (\ref{4}) normalized by its integrated
intensity for the chiral fluctuation operator $A_{\bf q}^{\chi}$ at
${\bf q} = (0,0)$ of the Hamiltonian (\ref{1}) at various values
of the coupling ratio $J_2/J_1$, obtained via strong-coupling
continued-fraction reconstruction from the coefficients
$\Delta _1,...,\Delta _6$ and a Gaussian terminator.}
\label{F4}
\end{figure}

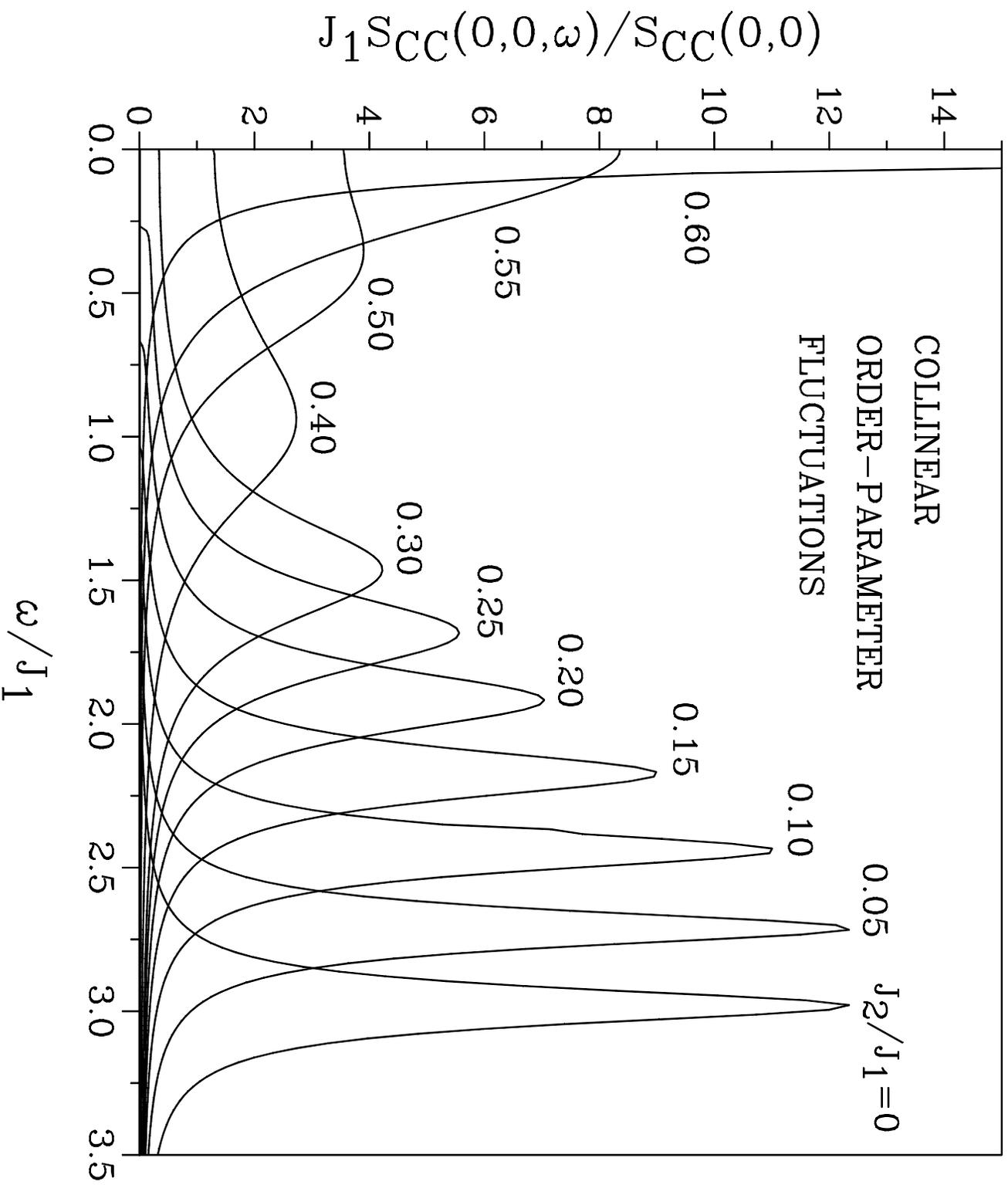
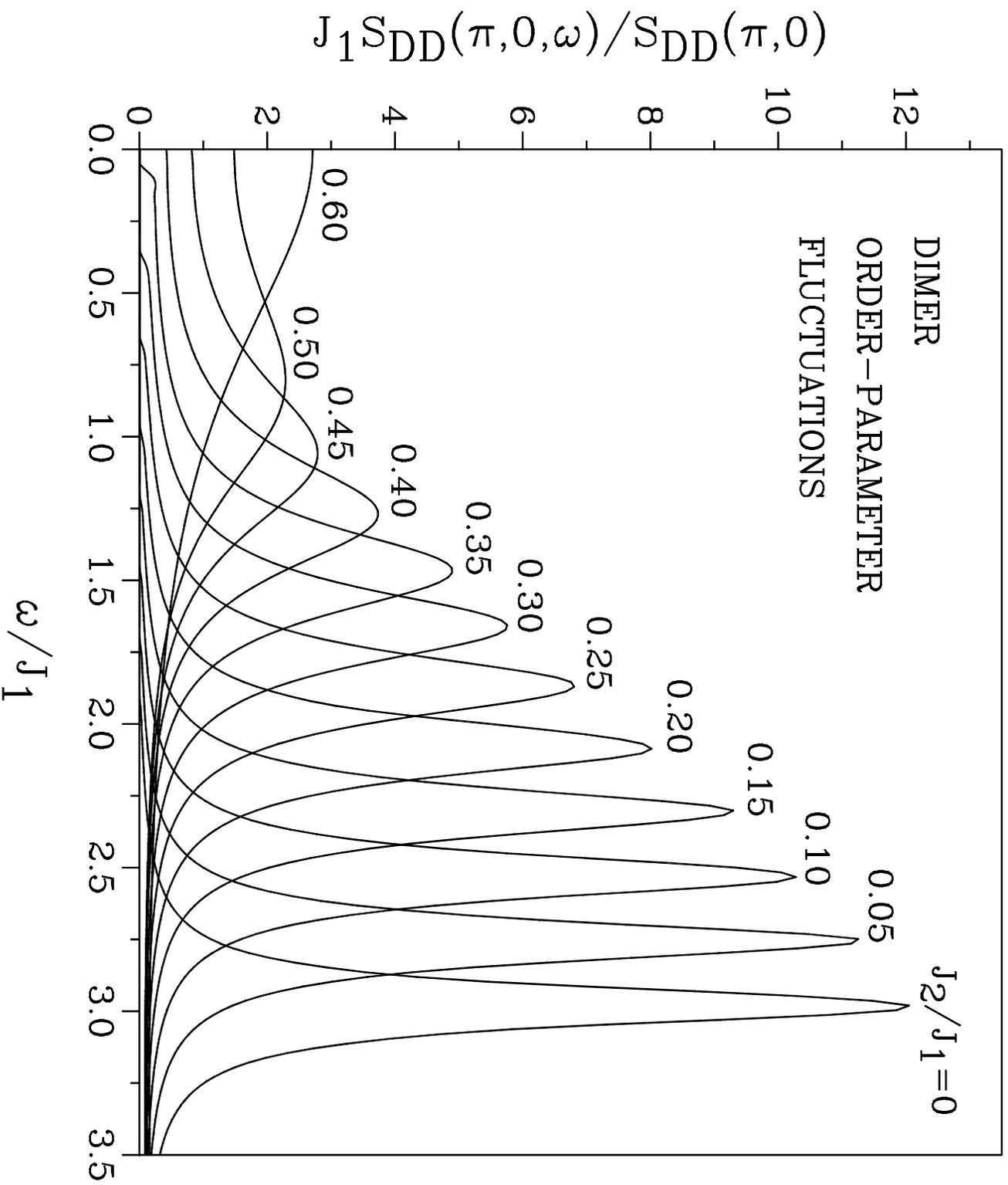
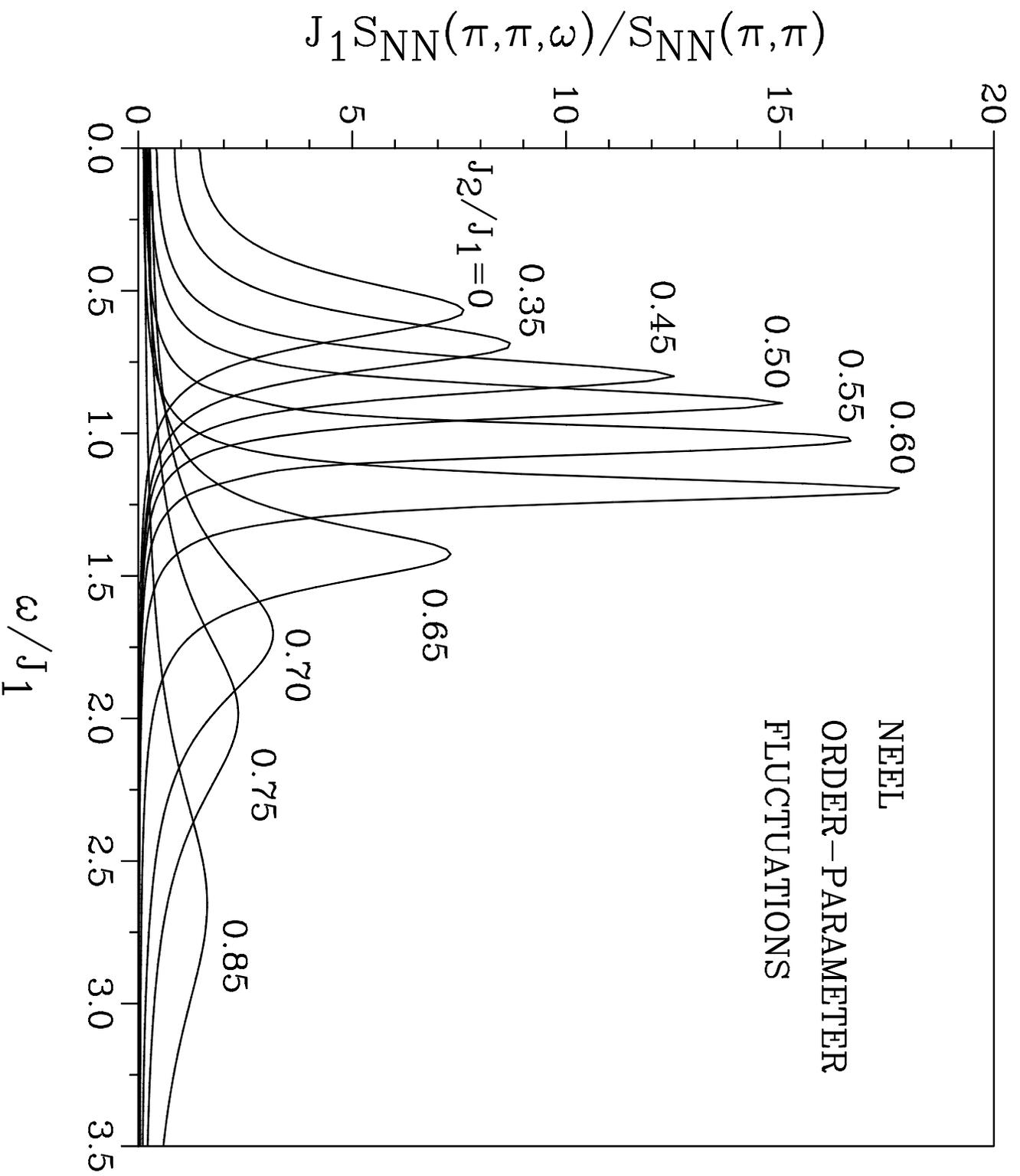
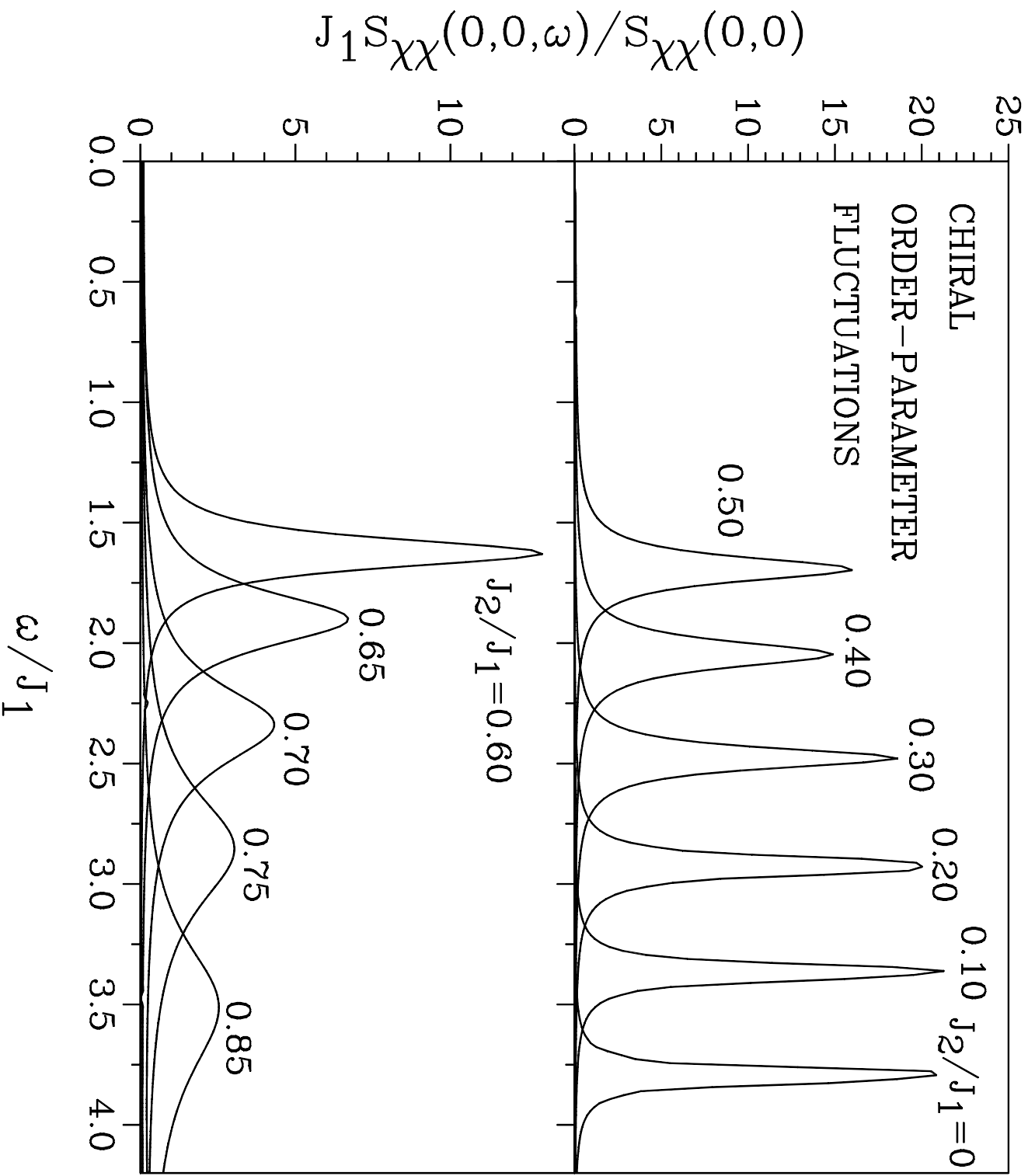
\end{document}